\documentclass[11pt,twoside]{article}
\usepackage{asp2004}
\usepackage{psfig}
\usepackage{epsf}
\begin{document}
\title{A new reddening for RR Tel from the HeII Paschen lines}
\author{Pierluigi Selvelli$^{1,2}$ and Piercarlo Bonifacio$^2$}
\affil{$^1$INAF-IASF, Roma,
$^2$INAF, Osservatorio Astronomico di Trieste,  Italy}

\begin{abstract}
From the available STIS data of RR Tel,  that have provided a
coverage with absolutely calibrated data in a wide wavelength range,  we
have obtained a new determination of its
reddening (E$_{B-V}$=0.00) from the comparison of the observed HeII
Paschen
lines decrement relative to HeII $\lambda$ 4686 (for 24 HeII Paschen lines
down
to the region of the head of the series near lambda 2060 \AA) with the
theoretical one as given in Storey and Hummer (1995) for case B, T=10,000
K and log N$_{e}$=6. This new  E$_{B-V}$=0.0 value has been confirmed from
a
re-analysis of the IUE low resolution data. We recall that  the so far
generally adopted value in the literature has been E$_{B-V}$=0.10 as
obtained by Penston et al. (1983). 
\end{abstract}

\section{The STIS  data}

We have retrieved from the HST archive  all STIS  spectra secured on
Oct. 10, 2000 by Keenan et al., in particular the  echelle   and the
grating spectra that cover the spectral regions of the HeII Paschen
lines. These STIS data have been recently
(Dec.
2003) re-calibrated  for the effects of increasing charge transfer
inefficiency and for the effects of time-dependent optical sensitivity.

\section{A new reddening and  distance }

Following Penston et al (1983) we have updated their method
of comparing the observed intensities of the He II recombination lines
with the theoretical ones in order to estimate  the reddening value. 
The STIS data provide a full coverage for the whole set of  HeII Paschen
lines
up to the region of the head of the series, near $\lambda$ 2050 \AA.

This  
spectral region of RR Tel has remained "unexplored" so far,  because 
 of the  low  response level
of the IUE LW cameras below lambda 2300 \AA. Instead, on STIS spectra one
can
resolve  the HeII Paschen lines up to the 37-3 transition at
2064.15 A, very near to the series limit  (see Fig. 1). 
The observed relative intensities (relative to I($\lambda$ 4686 \AA) =
100) have been
compared with
the theoretical ratios listed by Storey and Hummer (1995) for various log
Ne and Te values (case B). In Fig. 2  I$_{obs}$/I$_{th}$ is  plotted
against
wavelength  for theoretical values corresponding to log N$_{e}$=6 and
T$_{e}$=10,000 K. 

The points clearly define a straight line with slope =
0.00
(with some symmetric scatter of less than $10\%$ for the lines near the
head of the series), a direct evidence  that E$_{B-V}$=0.00.

\begin{figure}[!ht] 
\plotone{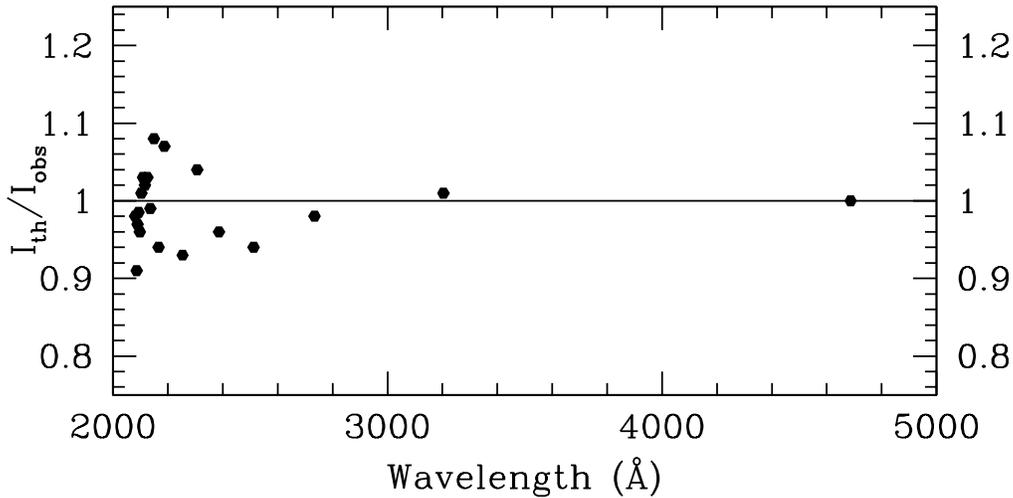}
\caption{The theoretical/observed HeII Paschen lines ratio  for  case B,
$\log$
$N_{e}$ = 6,  and $\log$ T = 4.  The emission  intensities are
normalized to $\lambda$ 4686. }
\end{figure}

We recall that Penston et al.  obtained E$_{B-V}$=0.10, but their
measurements were
based on a
limited number of He II lines and on more noisy  spectra. 
Penston et al. (1983) confirmed the  E$_{B-V}$=0.10 value obtained from
the
Paschen lines from  the alleged  presence of an absorption bump near
lambda
2175 in the continuum of IUE low resolution spectra.
In order to verify  this  point, we have created  an "average"
spectrum out of 39 SW and 35 LW IUE  low resolution spectra (Fig 3). In
this high S/N "average" spectrum there is no evidence of the absorption
dip reported by Penston et al (1983). Probably, the "lack" of strong
emission features near $\lambda$ 2200 has mimicked the presence of an
absorption bump.

  The photometric and spectral development of RR Tel during the outburst
phases
was that of an extremely slow nova with t$_3$ $\sim$ years. With the
assumption that the nova luminosity was near-Eddington
during these decay phases one obtains  M$_{bol}^{max}$$\sim$-6.1  and 
M$_{v}^{max}$$\sim$-6.0
(if  BC $\sim$
-0.1 for an object with T $\le$ 10000 K), in good
agreement  with  the estimates from various  MMRD relations. 
Thus, from  the observed m$_{v}^{max}$=6.7 and the new value for the
extinction 
( A$_{v}$=0.0 )  a  distance  of  3.47 kpc  is obtained.
This value is in  good agreement with that of  3.6 kpc  obtained by  Feast
et al (1983) on the assumption that a Mira is present in RR Tel.

\bibliographystyle{aa}

\begin{thebibliography}{}
\bibitem[Feast et al.(1983)]{1983MNRAS.202..951F} Feast, M.~W., Whitelock, 
P.~A., Catchpole, R.~M., Roberts, G., \& Carter, B.~S.\ 1983, \mnras, 202, 
951 
\bibitem[Penston et al.(1983)]{1983MNRAS.202..833P} Penston, M.~V.,
et al.\ 1983, \mnras, 202, 833 
\bibitem[Storey \& Hummer(1995)]{1995MNRAS.272...41S} Storey, P.~J.~\& 
Hummer, D.~G.\ 1995, \mnras, 272, 41 

\end{thebibliography}

\end{document}